# Kepler 16: A System of Potential Interest to Astrobiologists

Martin J. Heath

Ecospheres Project, 47 Tulsemere Road, London SE27 9EH, U.K., *ecospheresproject@hotmail.co.uk*.

Laurance R. Doyle

Carl Sagan Center, SETI Institute, 189 Bernardo Avenue CA 94043, USA, *ldoyle@seti.org*

Abstract

*We use the circumbinary planetary system Kepler-16b as an example to specify some considerations that may be of interest to astrobiologists regarding the dynamic nature of habitable zones around close double star systems.*

Kepler 16b is the first circumbinary planet to be confirmed by transit (Doyle *et al.*, 2011). This planet has a mass comparable to Saturn (although it is more dense), and so is very unlikely to host life. However, if we consider a hypothetical moon around this planet, we could consider aspects of habitability such a moon would experience. Such habitable moons around giant planets could, in principle, support life in certain cases. Jupiter's moon Europa and recently Saturn's moon Enceladus have been the focus of particular attention. Reynolds *et al.* (1983), Greenberg *et al.* (2000), Marion *et al.* (2003), McKay *et al.* (2008), Parkinson *et al.* (2008) and Rampellotto (2010) are among authors who have discussed the feasibility of life beneath the icy surfaces of these two bodies.

The characteristics of the Kepler-16 system have been detailed in Doyle *et al.*. The binary star system orbited by the planet Kepler-16b consists of a close (semi-major axis = 0.2243 AU, astronomical unit) pair of stars, one a K2-dwarf of 0.6897 $M_\odot$ (solar masses) and an M4.5-dwarf of mass 0.2026 $M_\odot$. The stars orbit each other every 41.07922 days with an orbital eccentricity of 0.1594. The planet occupies a nearly circular (e = 0.0069) 228.776-day orbit with a semi-major axis of 0.7048 AU. The K-star has an effective surface temperature of 4,450K and a radius of 0.6489 $R_\odot$ (solar radii). This corresponds to a luminosity of about 0.148 $L_\odot$ (solar luminosities).

The astronomical environment of Kepler 16b differs from that of the Solar System giant planets in that it receives significantly higher levels of insolation that the giant planet realm. Jupiter is the closest giant planet to the Sun, and the Jupiter system receives only ~ 0.04 $I_\oplus$ (Earth-level insolation) at perihelion, falling to ~0.03 $I_\oplus$ at aphelion. Kepler 16b, at a semi-major axis of 0.7048 ± 0.0011 AU, will receive, on average, an insolation of 0.298 $I_\oplus$. In the Solar System, for comparison, this insolation level would correspond to a distance from the Sun of 1.83 AU.

At first sight, the latter may appear tantalisingly close to the outer margin of the classic circumstellar habitable zone (HZ) of Kasting *et al.* (1993)—the zone constrained by climatic models in which a suitable planet or large moon could maintain liquid water at its surface.

There the estimated greenhouse limit (at which increasing $CO_2$ levels would fail to raise surface temperatures, due to the increased reflectivity of dense, and condensing, $CO_2$ atmospheres) would be located at about 1.67 AUs from the Sun (as a conservative limit). A not-so-conservative limit by Forget & Pierrehumbert (1997) later argued that with down-scattering of infrared light by clouds of $CO_2$ ice crystals—and for an atmosphere with 10 bars of $CO_2$—a planet could maintain at a surface temperature of > 0°C at up to 2.4 AU from the Sun (0.17 $I_\oplus$).

It would, however, be a mistake to directly compare insolation levels on possible moons with those for the HZ estimated using climatic models for Earth-like planets with substantial atmospheres. Key to these models was the operation of the natural thermostat of the carbonate silicate cycle (Walker *et al.*, 1981), which requires a significant level of geological activity on a terrestrial-type planet. We cannot, of course, entirely rule out large-terrestrial-planet type moons, although models by Canup & Ward (2006) explained the fact that Solar System satellites of a given planet should have maximum masses < $2.5 \times 10^{-4}$ of their primary in terms of the limited amount of material available to form moons in circum-planetary disks. Indeed, by reference to Solar System giant planet moons, which have a major icy component, we have no reason to assume that moons of giant planets elsewhere will have similar masses, compositions, geodynamics or atmospheric cyles comparable to that of the terrestrial planets. Notwithstanding, if the dynamical state of the Kepler 16b system is compatible with the survival of moons, then, with insolation values closer to those of present day Mars than those of Europa, astrobiologists may have cause to take note.

The architecture of the system will impose a non-trivial degree of seasonality even were moons of Kepler 16b to possess 0° obliquity with regard to the orbit of Kepler 16b around the binary. In estimating the cycle of insolation it is necessary to remember that the planet and both stars are orbiting not the larger star, but the barycentre of the system. This means that seasonality will occur even though, at the present era, the planet has a neglibile orbital eccentricity.

The distance of the stars from their barycentre will be proportional to their relative masses. At mean separation, the K-star will lie 0.050 AU and the M-star (considered here to provide negligible insolation) around 0.17 AU from the barycentre. The displacement of the K-star will decrease to 0.042 AU at periastron and increase to 0.058 AU at apoastron.

Minimum insolation will be received when the K-star lies at apoastron and on the opposite side of the barycentre to the planet. At such times it will lie 0.763 AU away and the planet will receive 0.254 $I_\oplus$ —over six times the greatest insolation at Jupiter's moon, Europa.

Maximum insolation will be received at those times when the K-star lies farthest from the system barycentre and when, simultaneously, the K-star – planet distance is smallest. At such times the K star will lie 0.058 AU outwards from the barycentre and the centre of the K star will lie at 0.646 AU from the planet, which will therefore receive > 0.354 $I_\oplus$. This may be compared with the 0.36 $I_\oplus$ received by Mars when it lies at aphelion.

The synodic period for the planet and the binary is 50.07 days, but this is not the interval for the recurrence of the above configurations, which would require that the planet be aligned along a projection of the binary line of apsides and that the K-star and the planet are either 180° apart or at the same barycentric longitude. Relevant cyclicities remain to be investigated, but we note that (ignoring as negligible any possible apsidal precession in the binary over the interval in question) ~ 39 orbits of the binary correspond to ~ 7 orbits of the planet, or nearly 4.4 Earth years.

In addition, a dynamical study (Doyle et al. 2011, supporting online material) has indicated that the planetary orbit will undergo cyclic changes in eccentricity up to a value of ~ 0.09 over a period of 2 Myr. If the dynamical evolution of the system were to permit the line of apsides of the binary and that of the planet to coincide, such that the planet could come to opposition at its closest point to the barycentre—when the K star were at binary apoastron—then the distance between the planet and barycentre would be 0.634 AU, and that to the centre of the K star would be reduced to a mere 0.583 AU. Under these conditions, insolation would rise to > 0.435 $I_\oplus$ which is comparable to that of Mars, 0.52 $I_\oplus$ at perihelion.

Kepler 16b does not, of course, initiate the discussion of whether extrasolar giant planets could possess habitable moons. Of 1,235 planet candidates identified in data from the Kepler mission, Borucki *et al.* (2011) noted 54 which might be deemed to lie in the HZ. Most were significantly larger than the Earth. Kasting (2011) discussed these candidates from a HZ perspective, noting that "dozens of gas giant planets within stellar habitable zones were already known from prior RV surveys. Life could still exist within these systems if these planets have large moons (Williams *et al.*, 1997); however, the prospects for detecting life on such moons are extremely slim." Such caution is appropriate, and the points made by Kasting (2011) should serve to curtail uncritical enthusiasm. Howbeit, without greeting Kepler 16b with a naïve fanfare, the fact remains that it does pose challenges for astrobiologists.

Moreover, the hypothesis that icy moons may harbour sub-surface life, whilst difficult to pursue in the case of Kepler 16b (which lies about 200 light years away), is open to being tested directly in the course of missions within the Solar System, in the asteroid belt, around giant planets, or in the Kepler Belt.

A number of authors have discussed how, as our Sun moves off the main sequence and enters its red giant phase, distant moons and dwarf planets rich in water and organics may become habitable. Lorenz *et al.* (1997) discussed the implications for Titan, and Stern (2003) noted what he dubbed the "Delayed Gratification Habitable Zone," which will eventually exist in the Kuiper Belt. In contrast, astronomers have sought to account for the presence of gas giant planets close to their stars in terms of inward migration from a zone of formation beyond the "snowline" (see, for example, review by Papaloizou & Terquem, 2005). The fates of icy moons carried into higher insolation regimes as their parent giant planets migrate starward, early in the history of planetary systems, provide an intriguing counterpoint to the later outward migration of Earth-like insolation levels as, late in their histories, stars evolve into highly luminous red giants.

The evolution of icy moons which have migrated inwards remains to be modelled in detail, but it is an issue of prime concern to astrobiology. In the case of a particular moon, it will depend upon many factors besides rising insolation levels. These will include the moon's mass, composition, structure, charged particle environment, impact history, tidal history and the history of its geodynamic regimes. Destinies of moons migrating inwards with giant planets may include becoming (particularly for the smallest icy moons) *de facto* comets which disintegrate leaving only dusty rings around their planets, becoming significantly massive silicate-dominated objects, retaining substantial shrouds of frozen volatiles, including high-pressure ices, or possessing subsurface oceans. The discovery of Kepler 16b may help to catalyse pursuit of appropriate models and rewarding thought experiments along these lines.